\documentclass[a4paper,11pt]{article}
\usepackage{pos}
\usepackage{hyperref}
\usepackage{mathrsfs}
\usepackage{amsmath}
\usepackage{epsfig}
\usepackage{epsf}
\usepackage{epsfig}
\usepackage{color}
\usepackage{xcolor}
\usepackage{ifpdf}
\usepackage{graphics}
\usepackage{longtable}
\usepackage{slashed}
\usepackage{float}
\restylefloat{table}
\usepackage{soul}
\usepackage{mathtools}

\usepackage[normalem]{ulem}

\newcommand{\ba}{\begin{eqnarray}}
\newcommand{\ea}{\end{eqnarray}}
\newcommand{\be}{\begin{equation}}
\newcommand{\ee}{\end{equation}}

\title{Nucleon Transversity from lattice QCD}

\author*{Constantia Alexandrou}

\affiliation{Department of Physics, University of Cyprus, PO Box 20537, CY-1678 Nicosia, Cyprus}

\emailAdd{alexand@ucy.ac.cy}

\abstract{We give a brief overview of recent progress in lattice QCD simulations that is enabling  precision studies of the three-dimensional  structure of the nucleon. We present results on  nucleon charges and second Mellin moments of parton distribution functions and generalized parton distributions, highlighting results obtained at the continuum limit  using only gauge ensembles  simulated with physical quark masses. The tensor charge and transversity moments are determined with controlled systematics and used to extract nucleon transverse density distributions.  We also discuss progress towards the direct evaluation of generalized parton distributions  in lattice QCD and the impact they are having on phenomenology. }

\FullConference{7th International Workshop on “Transverse phenomena in hard processes and the transverse structure of the proton (Transversity2024)\\
 3-7 June 2024\\
Trieste, Italy\\}

\begin{document}
\maketitle

\section{Introduction}\label{sec:intro}
The fundamental theory of the strong interaction, Quantum Chromodynamics (QCD), is known for  50 years~\cite{Fritzsch:1973pi}. Its Lagrangian has as input parameters the  six quark masses and the strong  coupling constant. Compared to the electroweak theory, it exhibits the unique properties of  confinement and asympotic freedom and generates 99\% of  the mass of visible matter in the universe. It describes the interactions in the microcosmos of sub-atomic particles but also governs the formation of very large objects like  neutron stars. Solving the theory is difficult due to its non-perturbative nature. The lattice formulation introduced by K. Wilson~\cite{Wilson:1974sk} provides a non-perturbative regularization of QCD  by introducing a momentum cut-off. The lattice formulation is well suited for a theory like QCD with asympotic freedom where the ultra violet behaviour is pertubative and the continuum limit well defined. Specifically, renormalization is implemented  by requiring  physical quantities to become independent of the lattice spacing as $a\rightarrow 0$. The formulation also provides the foundation for a numerical simulation since the finite lattice spacing and volume limits the number of degree of freedom.

\begin{figure}[h!]
\begin{minipage}{0.59\linewidth}
\includegraphics[width=\linewidth]{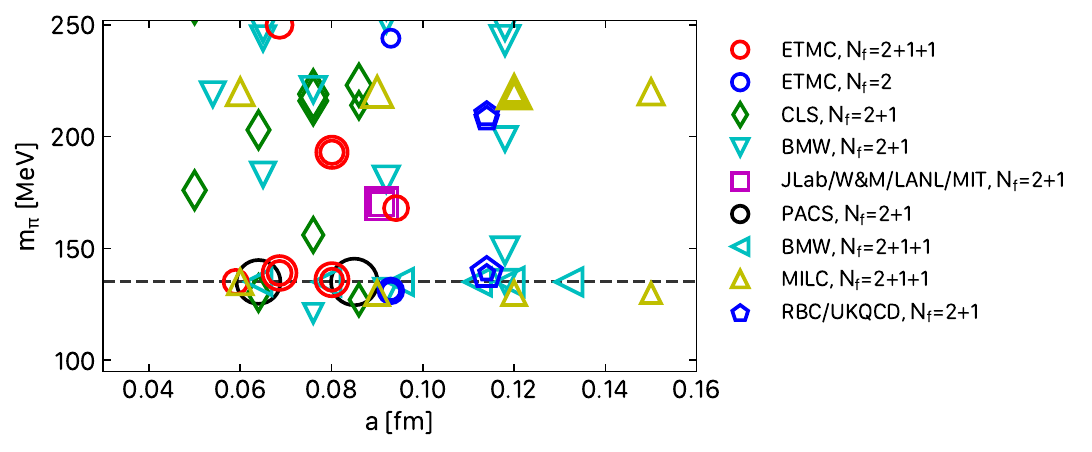}
\end{minipage}
\begin{minipage}{0.39\linewidth}
\includegraphics[width=\linewidth]{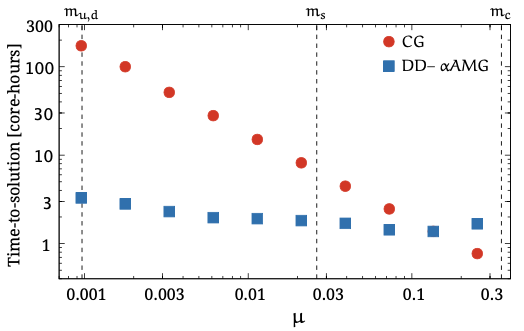}
\end{minipage}
\caption{Left: Gauge ensembles for different lattice spacings and pion masses. The size of the circles is proportional to the spatial volume of the lattice. The top 6 collaborations listed in the legend   use Wilson-type fermions,  the following two staggered fermions and the last one domain wall fermions. Right: Time to solution for inverting the fermion matrix $D_f$ as a function of the light quark mass, when using the multigrid (blue squares) as compare to the conjugate gradient (red circles). The horizontal dashed lines correspond, from right to left,  to the charm, strange and light (up and down) quark mass.}\label{fig:sims}
\end{figure}

Lattice QCD simulation utilizes the path integral representation of the theory in Euclidean space. The quantity of interest is expressed as the expectation value of an appropriate operator ${\cal {O}}$
\begin{equation} \langle {O}\rangle =\frac{1}{Z} \int {\cal D}[U] {\cal O}(D_f^{-1}[U], U) \left (\prod_{f} {\rm Det}(D_f[U]) \right)e^{-S_{\rm QCD}[U]} ,
\end{equation}
where $S_{\rm QCD}[U]$ is the gluon QCD action.
To compute  expectation values, one needs to generate an ensemble of gauge configurations $\{U\}$ with probability $ P[U] =\frac{1}{Z}  \left (\prod_f {\rm Det}(D_f[U]) \right)e^{-S_{\rm QCD}[U]}$, compute the inverse of the fermion matrix $D_f$ for every flavor $f$ or quark propagator and perform the specific contractions for the given operator ${\cal{O}}$. There are various lattice discretization schemes for the fermion part of the QCD action, each with its own merits and drawbacks. In actual simulations one includes the degenerate up and down (collectively called light quarks) and strange quarks. Such gauge ensembles are denoted as $N_f=2+1$ ensembles. In some simulations also  the charm  quark is added ($N_f=2+1+1$ ensembles). In the continuum limit, all discretization schemes with the same number of  flavors should give the same expectation value. In Fig.~\ref{fig:sims}, we compile the gauge ensembles currently available for the different discretization schemes that are have been produced by different collaborations. As can be seen, all major collaborations are producing gauge ensembles using the physical values of the light quark mass at various lattice spacings and volumes. In Fig.~\ref{fig:sims}, we also show the two orders of magnitude improvement at the light quark mass that resulted by the development of multigrid algorithms as compared to the previously used conjugate gradient for computing quark propagators. Such logarithmic improvements together with larger computational resources are enabling unprecedented accuracy in the determination of  a wealth of interesting quantities in lattice QCD.  For nucleon structure, Wilson-type discretization  schemes are particularly suitable and used for the results that followed. 
\section{Nucleon first and second Mellin moments}\label{sec:Mellin}
The three-dimensional (3D) structure  of the nucleon is encoded in the generalized parton distributions (GPDs) and transverse  momentum distributions  (TMDs). Here we will present results for the former. Since GPDs are defined on the light-cone,  until recently it was thought impossible to compute them directly in lattice QCD.   In the next section, we will discus the formalism that has become available to study GPDs  within lattice QCD, that can also be applied to TMDs. In this section, we present results on moments of GPDs for which lattice QCD calculations have a long history and where we can now reach unprecedented  accuracy. To extract such moments,   light-cone operators  are expanded in terms of a tower of local twist-2 operators the matrix elements of which  can be evaluated in lattice QCD. Following the notation of Ref.~\cite{Hagler:2009ni}, the  moments of the three types of PDFs are determined by computing the nucleon matrix elements of the following operators
 \begin{eqnarray}
      {\cal O}^{\mu\mu_1\ldots\mu_{n-1}}      = \bar{\psi}  \gamma^{\{\mu}iD^{\mu_1}\ldots iD^{\mu_{n-1}\}} \psi &\stackrel{unpolarized}{\rightarrow}&
\langle x^{n-1}\rangle_q = \int_{0}^{1}dx \, x^{n-1}\left[q(x)-(-1)^{n-1}\bar{q}(x)\right] \> \\
\tilde{\cal O}^{\mu\mu_1\ldots\mu_{n-1} }     =\bar{\psi} \gamma_5\gamma^{\{\mu}iD^{\mu_1}\ldots iD^{\mu_{n-1}\}} \psi & \stackrel{helicity}{\rightarrow}&
\langle x^{n-1}\rangle_{\Delta q} = \int_{0}^{1}dx \, x^{n-1}\left[\Delta q(x)+(-1)^{n-1}\Delta\bar{q}(x)\right]  \nonumber \\
      {\cal O}^{\rho\mu,\mu_1\ldots\mu_{n-1}}_T = \bar{\psi}  \sigma^{\rho\{\mu}iD^{\mu_1}\ldots iD^{\mu_{n-1}\}} \psi &\stackrel{transversity}{\rightarrow} &\langle x^{n-1}\rangle_{\delta q}=\int_{0}^{1}dx\, x^{n-1}\left[\delta q(x)-(-1)^{n-1}\delta\bar{q}(x)\right], \nonumber
   \end{eqnarray}
where
$ q=q_\downarrow+q_\uparrow$, \,\, $\Delta q=q_\downarrow-q_\uparrow$, \,\,$\delta q=q_\intercal+q_\perp $. 
For off-diagonal nucleon matrix elements, we obtain moments of GPDs or Lorentz invariant  generalised form factors (GFFs).   E.g. for the  unpolarized case we have 
 \begin{eqnarray}
 \int_{-1}^1 dx\, x^{n-1} H(x,\xi,Q^2)&=&\sum_{i=0,2,\cdots}^{n-1} (2\xi)^i A_{ni}(Q^2)+(2\xi)^{n} C_{n0}(Q^2)|_{n \, \rm{even}}\nonumber\\
\int_{-1}^1 dx\, x^{n-1} E(x,\xi,Q^2)&=& \sum_{i=0,2,\cdots}^{n-1} (2\xi)^i B_{ni}(Q^2)-(2\xi)^{n} C_{n0}(Q^2)|_{n \, \rm{even}},
 \end{eqnarray}
 where $Q^2$  is the squared momentum transfer. In practice, mixing limits the order of moments that can be extracted to $n=4$  or to third-derivative operators. Familiar cases are  the first and second moments which we will discuss in detail. The  first moments yield the vector, axial-vector and tensor  nucleon charges and form factors. The second Mellin moments of PDFs give important information on the momentum  fraction $\langle x\rangle_q=A_{20}(0)$ and total angular momentum  $J_q=\frac{1}{2}[A_{20}(0)+B_{20}(0)]$ carried by each quark flavor in the nucleon. Among the first and second moments, the isovector axial charge is very well measured and has served for years as a benchmark quantity for lattice QCD computations. It is only recently, however, that one can compute it using only gauge ensembles generated with the light, strange and charm quark mass set to their physical value (referred to as physical point).
 \begin{figure}[h!]
\begin{minipage}{0.49\linewidth}
\includegraphics[scale=0.35]{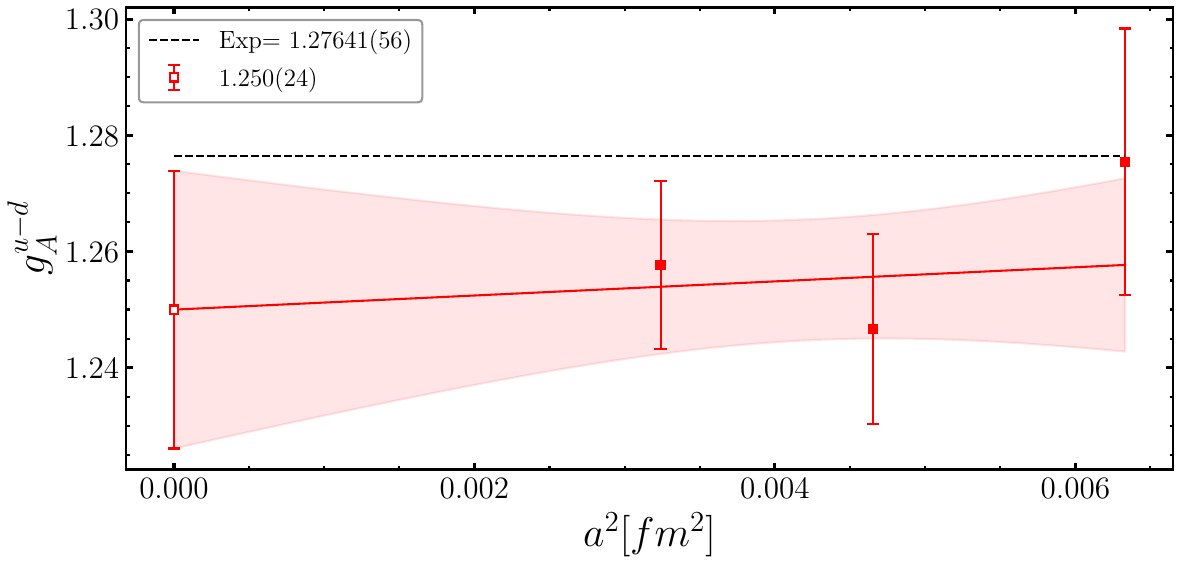}
\end{minipage}
\begin{minipage}{0.49\linewidth}
\includegraphics[scale=0.5]{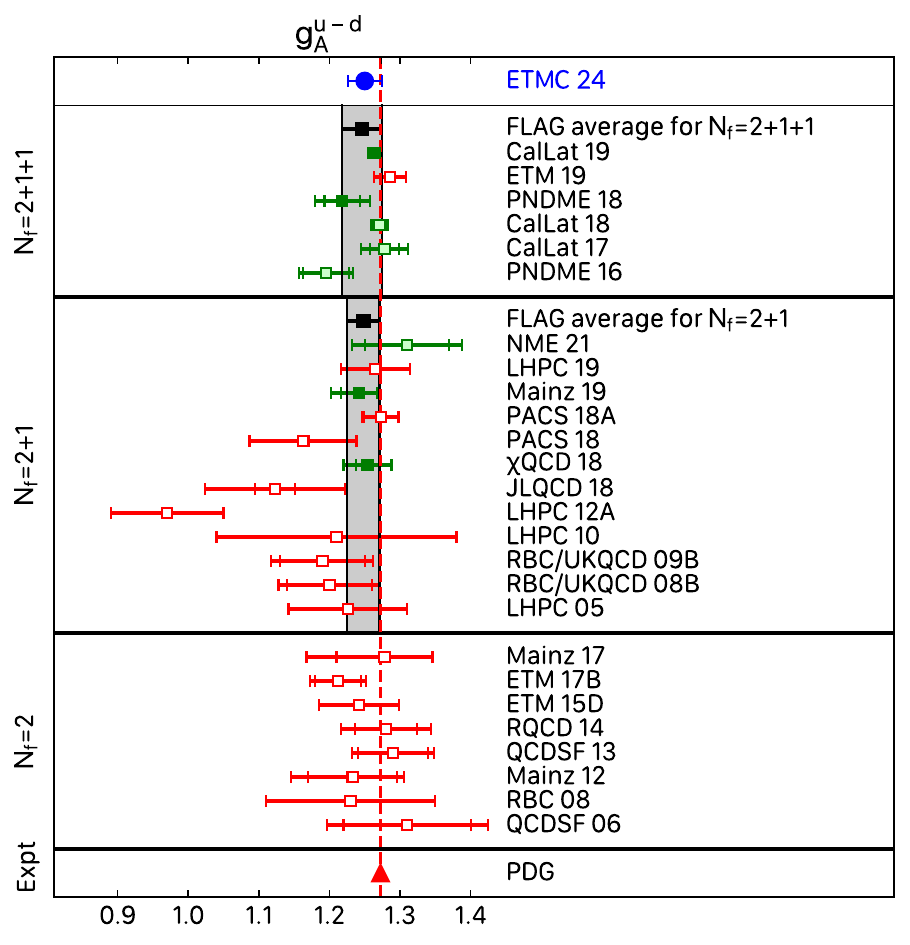}
\end{minipage}
\caption{Left: The isovector nucleon axial charge computed using three $N_f=2+1+1$ physical point twisted mass fermion ensembles. The open symbol is the continuum extrapolation result. Right: A compilation of the isovector axial charge by  FLAG2021~\cite{FlavourLatticeAveragingGroupFLAG:2021npn}. We  include the continuum result shown in the left panel (blue circle). The experimental value is shown by the red dashed line. The black squares show the  FLAG averages.
}\label{fig:axial}
\end{figure}

\begin{figure}[h!]
\begin{minipage}{0.49\linewidth}
\includegraphics[scale=0.35]{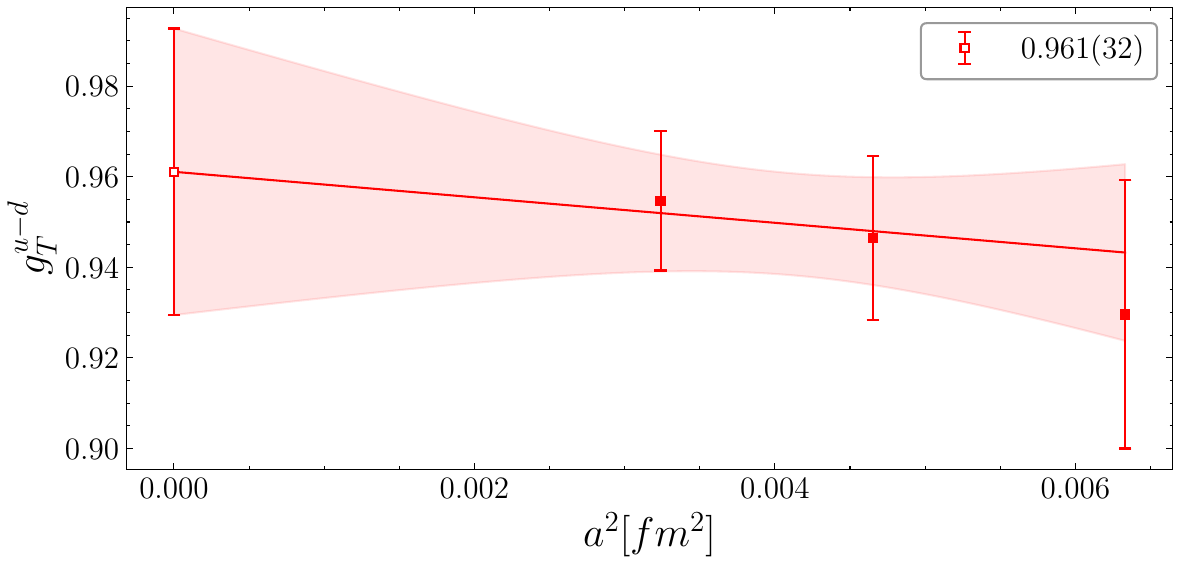}
\end{minipage}
\begin{minipage}{0.49\linewidth}
\includegraphics[scale=0.5]{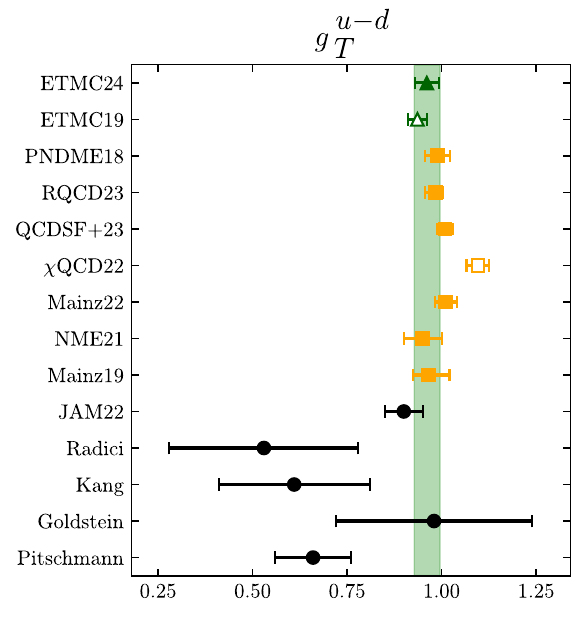}
\end{minipage}
\caption{Left: The isovector nucleon tensor charge in the $\overline{\rm MS}$ scheme at a scale of 2 GeV computed using three $N_f=2+1+1$ physical point twisted mass fermion ensembles. The open symbol is the continuum extrapolated result. Right: Comparison of ETMC results (green triangles) with other lattice QCD results (yellow squares)  and phenomenological extractions (black circles). Open symbols indicate lattice QCD results without continuum extrapolation. }
\label{fig:tensor}\vspace*{-0.3cm}
\end{figure}

\begin{figure}[h!]
\begin{minipage}{0.3\linewidth}
\includegraphics[width=\linewidth]{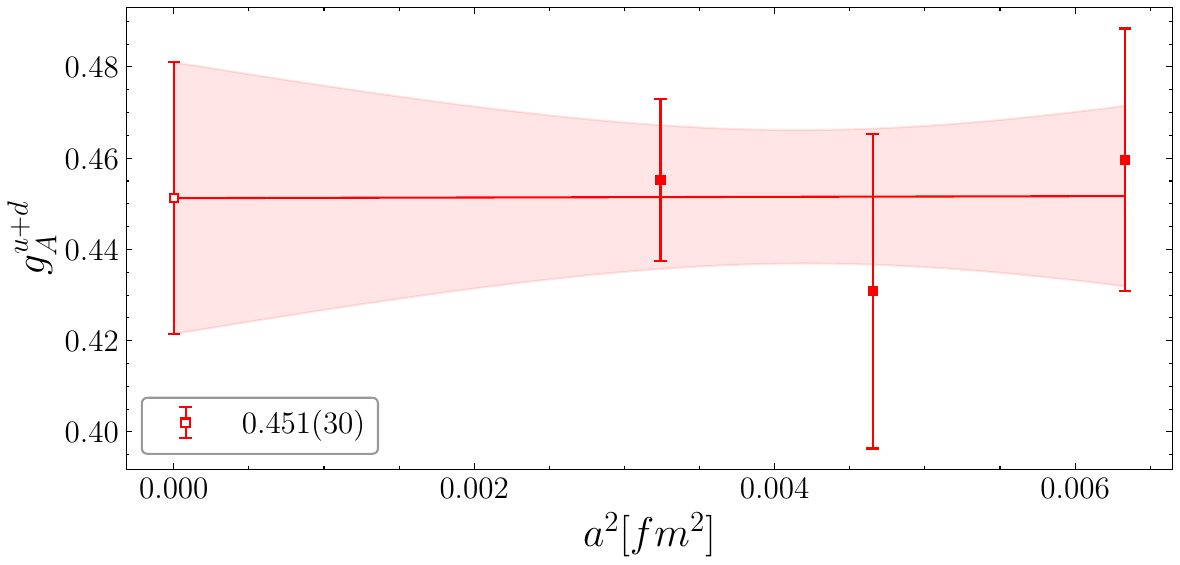}
\end{minipage}
\begin{minipage}{0.3\linewidth}
\includegraphics[width=\linewidth]{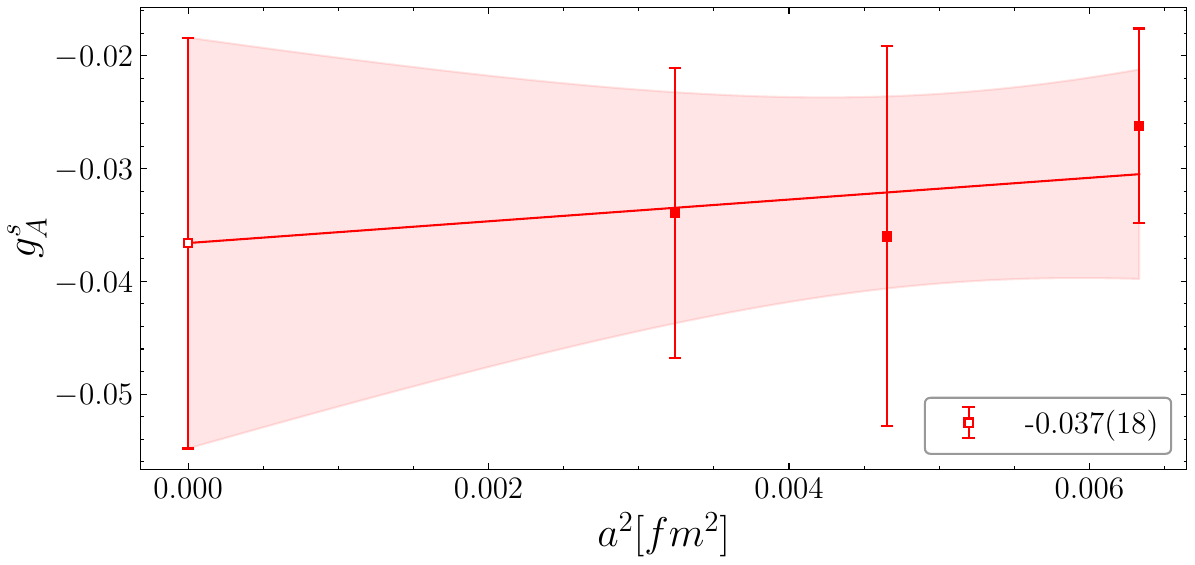}
\end{minipage}
\begin{minipage}{0.3\linewidth}
\includegraphics[width=\linewidth]{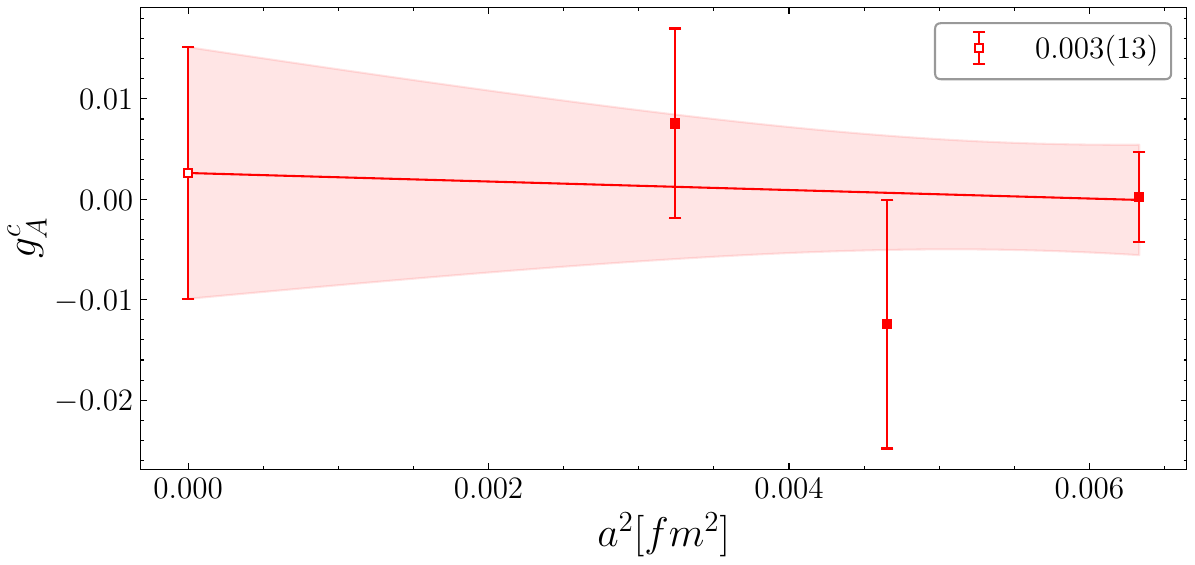}
\end{minipage}
\caption{The isoscalar (left), strange (middle)  and charm (right) axial charges for the nucleon. In the legend of each plot we give the value at the continuum limit.}\label{fig:intri}\vspace*{-0.3cm}
\end{figure}

In Fig.~\ref{fig:axial}, we show the continuum extrapolation  of the isovector nucleon axial charge $g_A^{u-d}$ using only physical point ensembles generated by the Extended Twisted Mass Collaboration (ETMC) and compare it with the compilation of lattice QCD results done by FLAG2021. As can been, lattice QCD results reproduce the experimental value. Postdiction of $g_A^{u-d}$ provides a validation of the lattice QCD approach that can be reliably now applied to determine less well measured quantities, like the isovector nucleon tensor charge $g_T^{u-d}$  for which results are shown in Fig.~\ref{fig:tensor}.  The lattice QCD  value of $g_T^{u-d}$ can be used to constrain  phenomenological extractions of  tensor PDFs and GPDs~\cite{Cocuzza:2023oam}.
Beyond the isovector charges that are technically easier to compute, one can also compute the charges for each quark flavor. Results for  the axial charges, including the continuum limit, are shown  in Fig.~\ref{fig:intri}. These determine the  intrinsic  spin $\Delta \Sigma_q$  carried by  quarks   in the nucleon. The corresponding values for the tensor charges in the $\overline{\rm MS}$ scheme at 2~GeV  are: $g_T^u=0.763(32)$, $g_T^d=-0.199(21)$, $g_T^s=-0.020(13)$ and $g_T^c=-0.0042(34)$.

\begin{figure}[h!]
\begin{minipage}{0.3\linewidth}
\includegraphics[width=\linewidth]{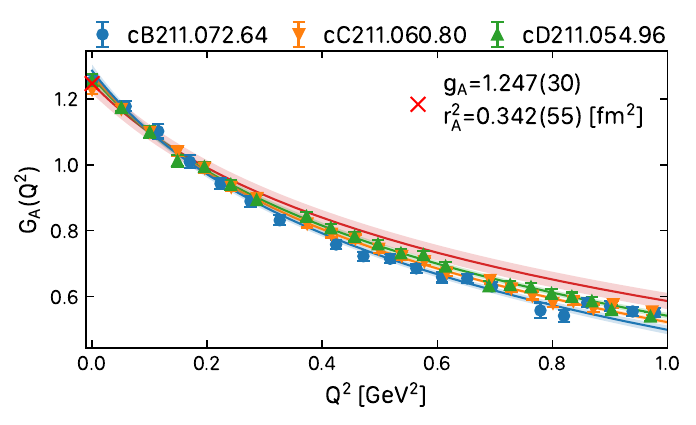}
\end{minipage}
\begin{minipage}{0.3\linewidth}
\includegraphics[width=\linewidth]{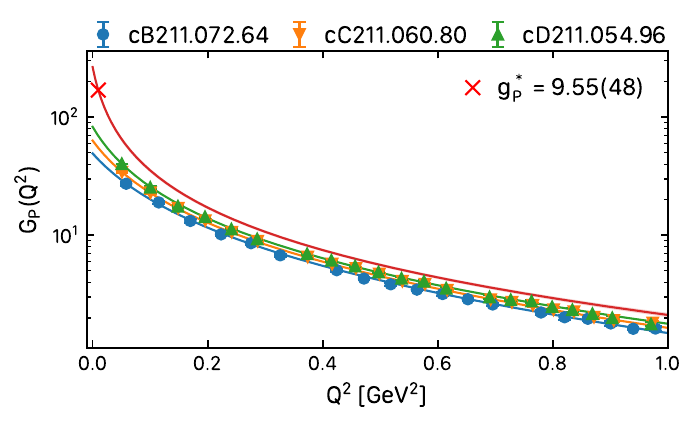}
\end{minipage}
\begin{minipage}{0.35\linewidth}
\includegraphics[width=\linewidth]{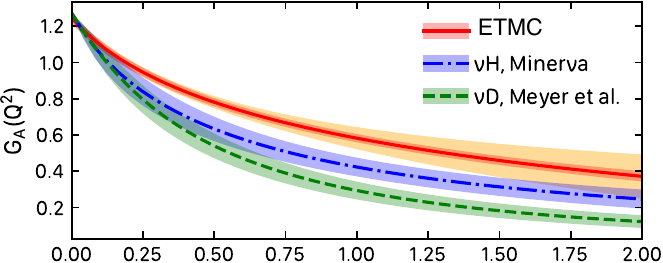}
\end{minipage}
\caption{The isovector axial FF $G_A(Q^2)$ (left) and induced pseudoscalar FF $G_P(Q^2)$  (middle) versus the momentum transfer squared $Q^2$, computed using three $N_f=2+1+1$ physical point twisted mass fermion ensembles. The red line  is the continuum extrapolation with the associated error shown by the red band. Right: Comparison  of $G_A(Q^2)$ from ETMC with experimental extractions with the blue band showing the latest results. The yellow band includes the systematic errors of the lattice QCD analysis by ETMC. }
\label{fig:axial FFs}
\end{figure}

Allowing momentum transfer  leads to the extraction of nucleon form factors (FFs). While the electromagnetic FFs are well measured, the axial FFs are not. In Fig.~\ref{fig:axial FFs}, we show ETMC results on the isovector axial and induced pseudoscalar FFs~\cite{Alexandrou:2023qbg}. The axial FF $G_A(Q^2)$ is an important input in neutrino experiments. As can be seen, more recent experimental data by the Minerva experiment are closer to the lattice QCD results.

\begin{figure}[h!]
\begin{minipage}{0.49\linewidth}
\includegraphics[width=\linewidth]{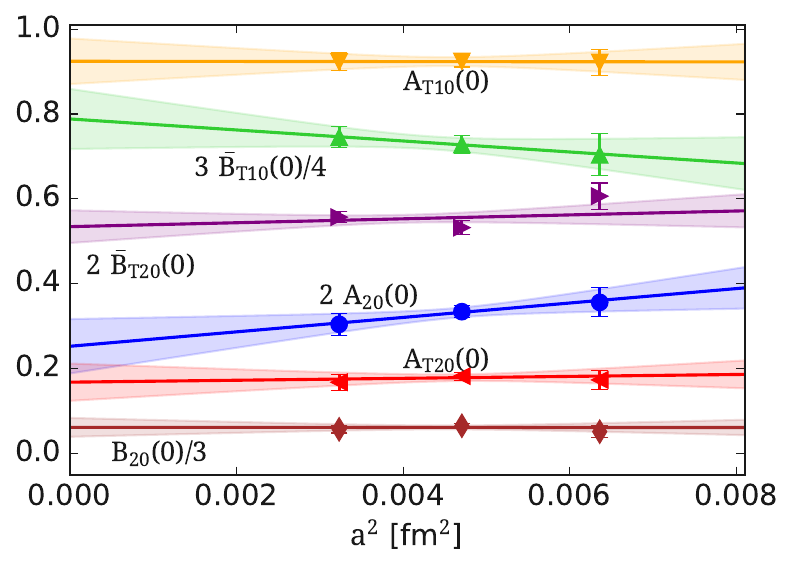}
\end{minipage}
\begin{minipage}{0.49\linewidth}
\includegraphics[width=\linewidth]{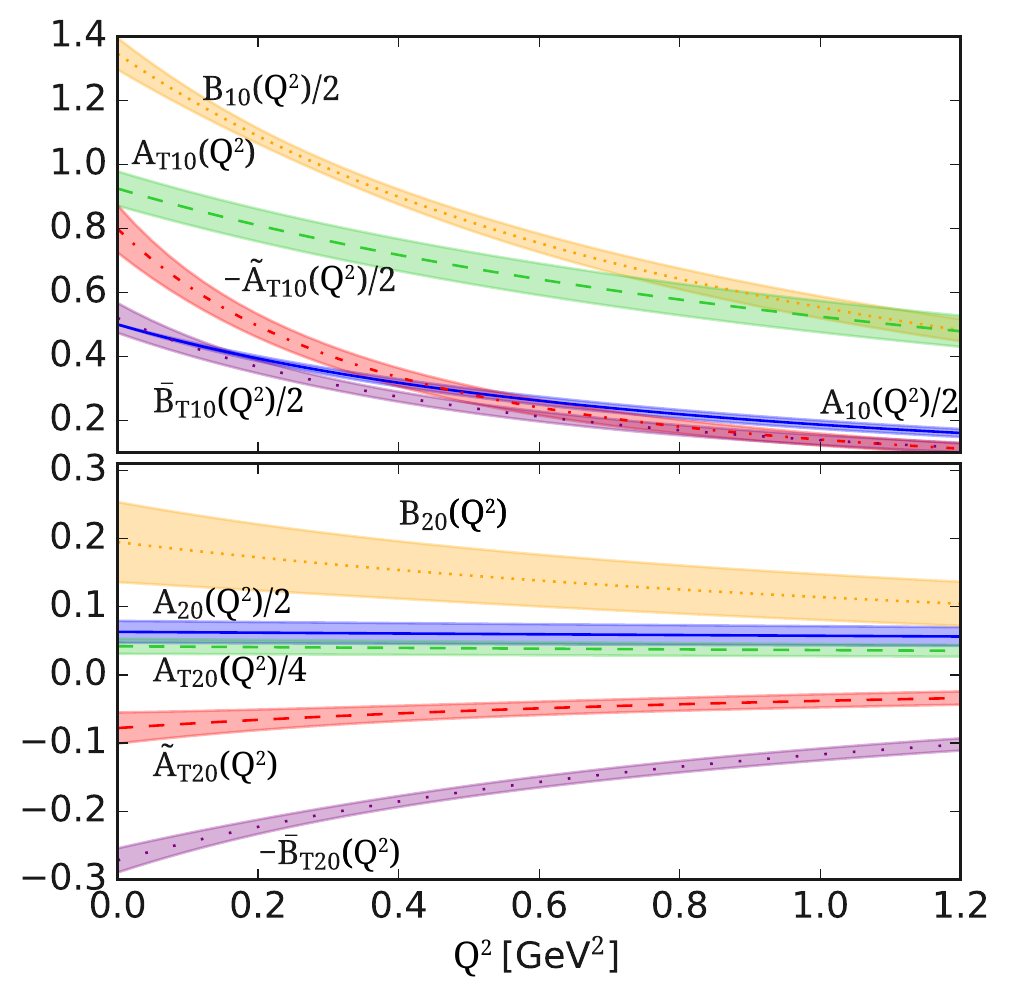}
\end{minipage}
\caption{Left: Continuum extrapolation of the isovector unpolarised and transversity GFFs for $Q^2=0$. Right: Fits after continuum extrapolation to the form $F(Q^2)=\frac{F(0)}{\left(1+Q^2/m^2\right)^p}$.}
\label{fig:GFFs}
\end{figure}

In what follows, we discuss the isovector spin densities in the transverse plane~\cite{Diehl:2005jf}. To determine them we need the unpolarized GFFs  of Eq.~3 extracted from
\be
\small
\langle N(p',s') \vert \mathcal{O}^{\mu\nu,f} \vert N(p,s) \rangle = \bar{u}_N(p',s') \Big[ A^{f}_{20}(Q^2) \gamma^{\{\mu} P^{\nu\}} + B^{f}_{20}(Q^2) \frac{i\sigma^{\{\mu\alpha} q_\alpha P^{\nu\}}}{2m_N} + C^{f}_{20}(Q^2) \frac{q^{\{\mu} q^{\nu\}}}{m_N} \Big] u_N(p,s) \ee
 and the first and second moments of the transversity GPDs extracted, respectively, from
\be
\small
\langle N (p^\prime,s^\prime)| \bar{\psi}_f i\sigma^{\mu\nu} \psi_f| N (p,s)\rangle =
\bar{u}_N(p^\prime,s^\prime)\left[i\sigma^{\mu\nu}\, A^f_{T10}(Q^2)
+ \frac{\gamma^{[\mu}q^{\nu]}} {2 m_N} \right. \,B^f_{T10}(Q^2)
+\left. \frac{P^{[\mu} q^{\nu]}} {m_N^2 }\, \widetilde A^f_{T10}(Q^2) \right]u_N(p,s)
\ee
and
\begin{eqnarray}
\small
   \langle N(p^\prime,s^\prime)| \mathcal{O}_{T}^{\mu\nu\rho,f}| N(p,s)\rangle &=&
    \bar u_N(p^\prime,s^\prime)\Bigl[A^f_{T20}(Q^2)\, i\sigma^{[\mu\{\nu]}P^{\rho\}} + \tilde A^f_{T20}(Q^2)\,\frac{P^{[\mu}q^{\{\nu]}P^{\rho\}}}{m_N^2} + \nonumber\\
   & & B^f_{T20}(Q^2)\, \frac{\gamma^{[\mu} q^{\{\nu]}P^{\rho\}}}{2m_N} + \tilde B^f_{T20}(Q^2)\,\frac{\gamma^{[\mu} P^{\{\nu]}q^{\rho\}}}{m_N}\Bigr]u_N(p,s). 
\end{eqnarray}
By the curl brackets we denote symmetrization over the Dirac indices and subtraction of the trace and with the square brackets antisymmetrization.  In Fig.~\ref{fig:GFFs}, we show the isovector  unpolarised  and   transversity GFFs, defined in Eqs. 4, 5 and 6, where the superscript $u-d$ is dropped. We use three physical point ensembles generated by ETMC to extrapolate to the continuum limit for each $Q^2$ value, as shown for the example of  $Q^2=0$ that determine the corresponding moments of PDFs. Instead of $B_{T10}$ and $B_{T20}$ it is customary to use  the combination $\bar{B}_{T10}=B_{T10}+2\widetilde{A}_{T10}$ and  $\bar{B}_{T10}=B_{T20}+2\widetilde{A}_{T20}$, respectively. Our values for the moments are as follows: i) The isovector anomalous tensor magnetic moment $\kappa_T=\bar{B}_{T10}(0)=1.051(94)$, which implies a non-zero Boer-Mulders function $ -h_1^\perp$; ii) the isovector momentum fraction $\langle x\rangle _{u-d}=A_{20}(0)=0.126(32)$ and total angular momentum $J_{u-d}=[A_{20}(0)+B_{20}(0)]=0.156(46)$; and iii) the isovector transversity moments $\langle x\rangle_{\delta u-\delta d}=A_{T20}(0)=0.168(44)$ and $\bar{B}_{T20}(0)=0.267(19)$~\cite{Alexandrou:2022dtc}.

\begin{figure}[h!]\vspace*{-0.6cm}
\begin{minipage}{0.49\linewidth}
\includegraphics[width=\linewidth]{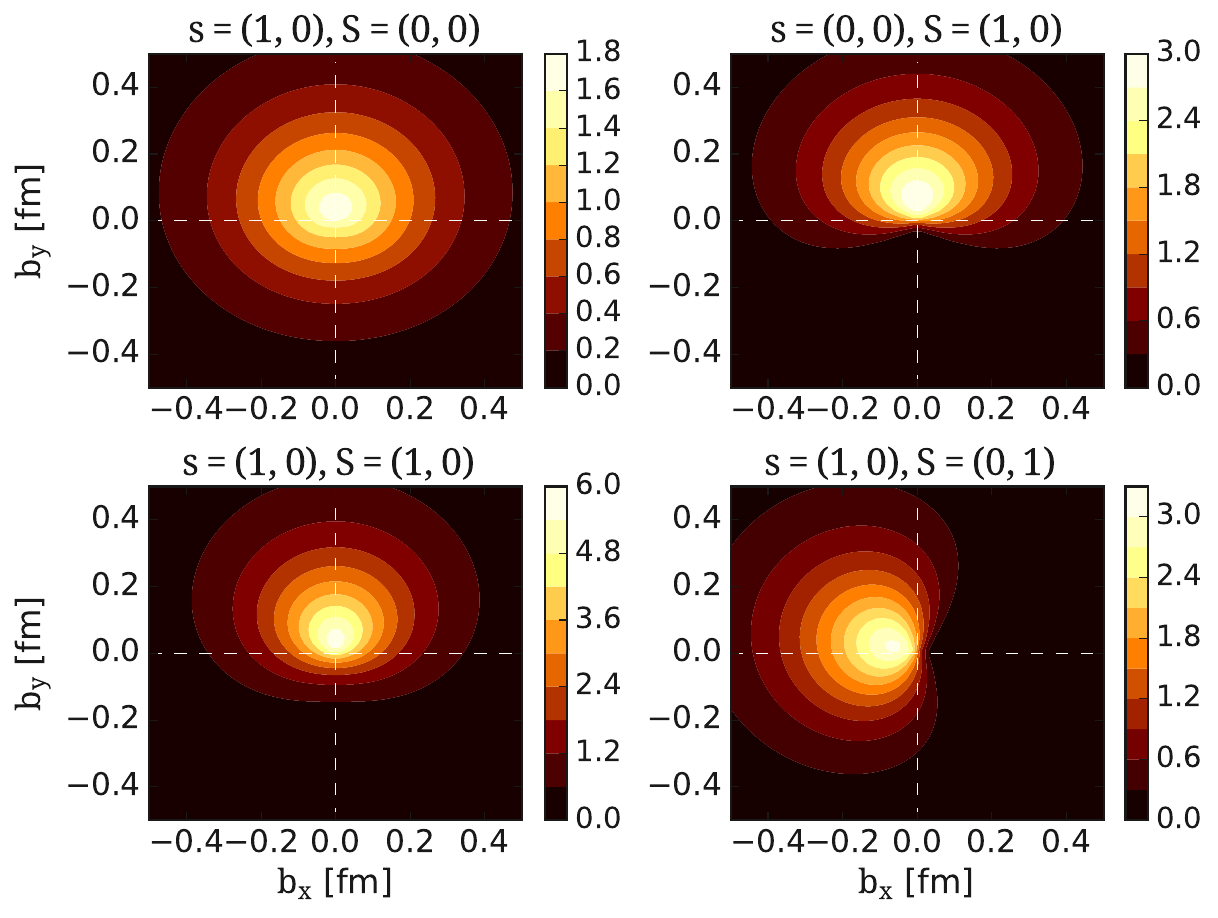}
\end{minipage}
\begin{minipage}{0.49\linewidth}
\includegraphics[width=\linewidth]{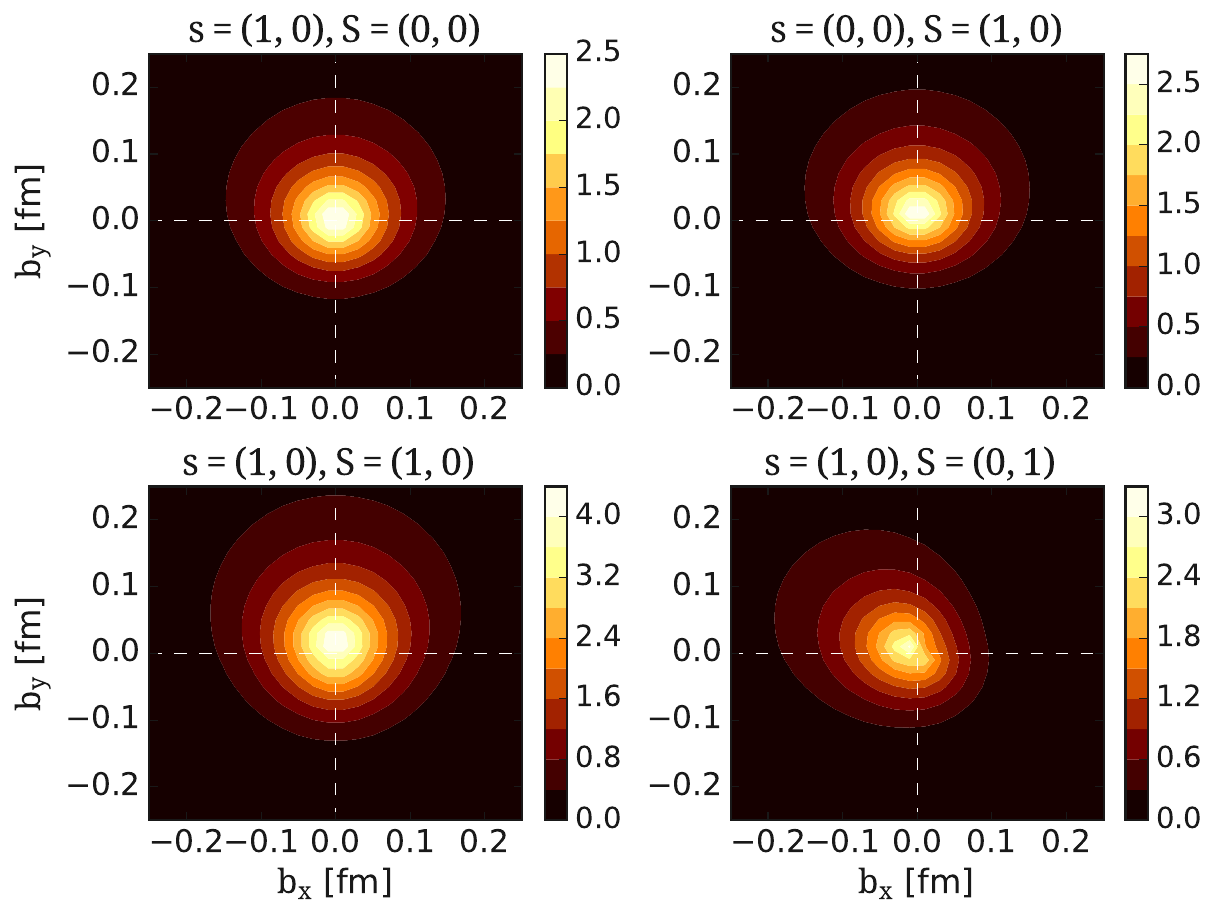}
\end{minipage}
\caption{Contours of the density distribution for the first  (left) and second (right) moments as a function of $b_x$ and $b_y$. For each moment we show: Upper left for transversely polarized quarks in an unpolarized nucleon; Upper right for unpolarized quarks in a transversely polarized nucleon; Lower left for  transversely polarized quarks in
a transversely polarized nucleon; and lower right for transversely polarized quarks in
a perpendicularly polarized nucleon.}\label{fig:rho}\vspace*{-0.5cm}
\end{figure}
After fitting the $Q^2$-dependence of the continuum extrapolated GFFs using the Ansatz $\frac{F(0)}{\left(1+Q^2/m^2\right)^p}$, we  Fourier transform the fitted Ansatz to impact parameter space. At zero skewness the spin densities in the transverse plane are given  by~\cite{Diehl:2005jf}
\begin{eqnarray}
    \rho(x,{\bf{b}}_\perp,{\bf{s}}_\perp,{\bf{S}}_\perp) &=& \frac{1}{2} \bigg[ H(x,b^2_\perp) +        \frac{{b}_\perp^j \epsilon^{ji}}{m_N} \left( {S}_\perp^i E'(x,b^2_\perp) + {s}_\perp^i \bar{E}'_{T}(x,b^2_\perp)\right) \\ &+&{s}_\perp^i {S}_\perp^i \left( H_T(x,b^2_\perp) - \frac{\Delta_{b_\perp} \tilde{H}_T(x,b^2_\perp)}{4 m_N^2} \right) 
    +{s}_\perp^i(2{b}_\perp^i {b}_\perp^j- \delta^{ij} b^2_\perp){S}_\perp^j \frac{\tilde{H}''_T(x,b^2_\perp)}{m_N^2} \bigg],\nonumber
\end{eqnarray}
where $F^\prime \equiv \frac{\partial}{\partial b_\perp^2}F,\,\,\, \Delta_{b_\perp}F\equiv 4\frac{\partial}{\partial b_\perp^2}(b^2_\perp \frac{\partial}{\partial b_\perp^2}) F$, $x$ is the longitudinal momentum fraction, $\bf{s}_\perp$ is the transverse quark spin, $\bf{S}_\perp$ is the transverse nucleon spin, and $\bf{b}_\perp$ is the transverse impact parameter. Taking moments $  \langle x^{n-1} \rangle_\rho ({\bf{b}}_\perp,{\bf{s}}_\perp,{\bf{S}}_\perp) \equiv \int_{-1}^{1} dx ~ x^{n-1} \rho(x,\bf{b}_\perp,\bf{s}_\perp,\bf{S}_\perp)$,
we obtain the transverse densities in terms of GFFs. The transverse density distributions for the two lowest moments,  $n=1$ and $n=2$, are shown in Fig.~\ref{fig:rho}. Qualitatively the behavior is similar to that  predicted in Ref.~\cite{Diehl:2005jf}. Namely, for the case $n=1$, we find a large asymmetry in particular when the nucleon is polarized. For the case of  unpolarized quarks in a transversely polarized nucleon, the  large effect is  due to the  $Q^2$ behavior of  $B_{10}$. The origin of this behavior is related to the Sivers
effect . The distortion for $n=2$ is milder than for $n=1$ due to the milder $Q^2$-dependence of $A_{20}$ compared to $A_{10}$.\vspace*{-0.3cm}

\section{Direct computation of PDFs and GPDs}\label{sec:quasi}
GDPs are light-cone correlation matrix elements and as such cannot be computed on a Euclidean lattice. However, X. Ji~\cite{Ji:2013dva} showed that one can relate spatial correlators for sufficiently large  boosted nucleons to the light-cone  GPDs using a perturbatively computed matching kernel within the large momentum effective theory (LaMET). 
\begin{figure}[h!]
\begin{minipage}{0.39\linewidth}
\includegraphics[height=0.7\linewidth]{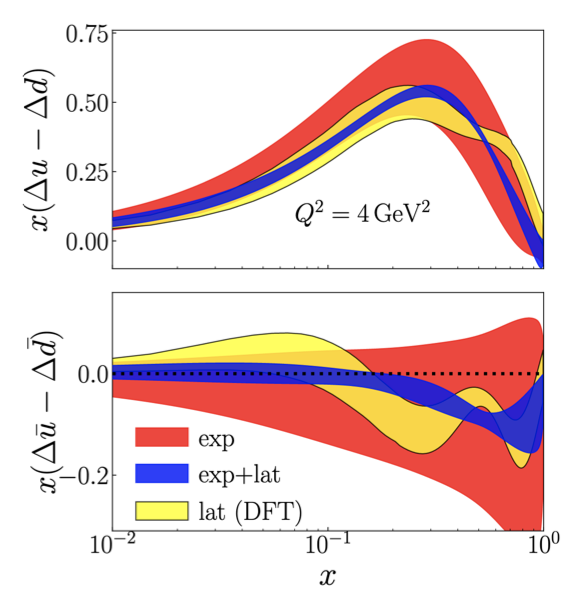}
\end{minipage}
\begin{minipage}{0.3\linewidth}
\hspace*{-2cm}\includegraphics[scale=0.23]{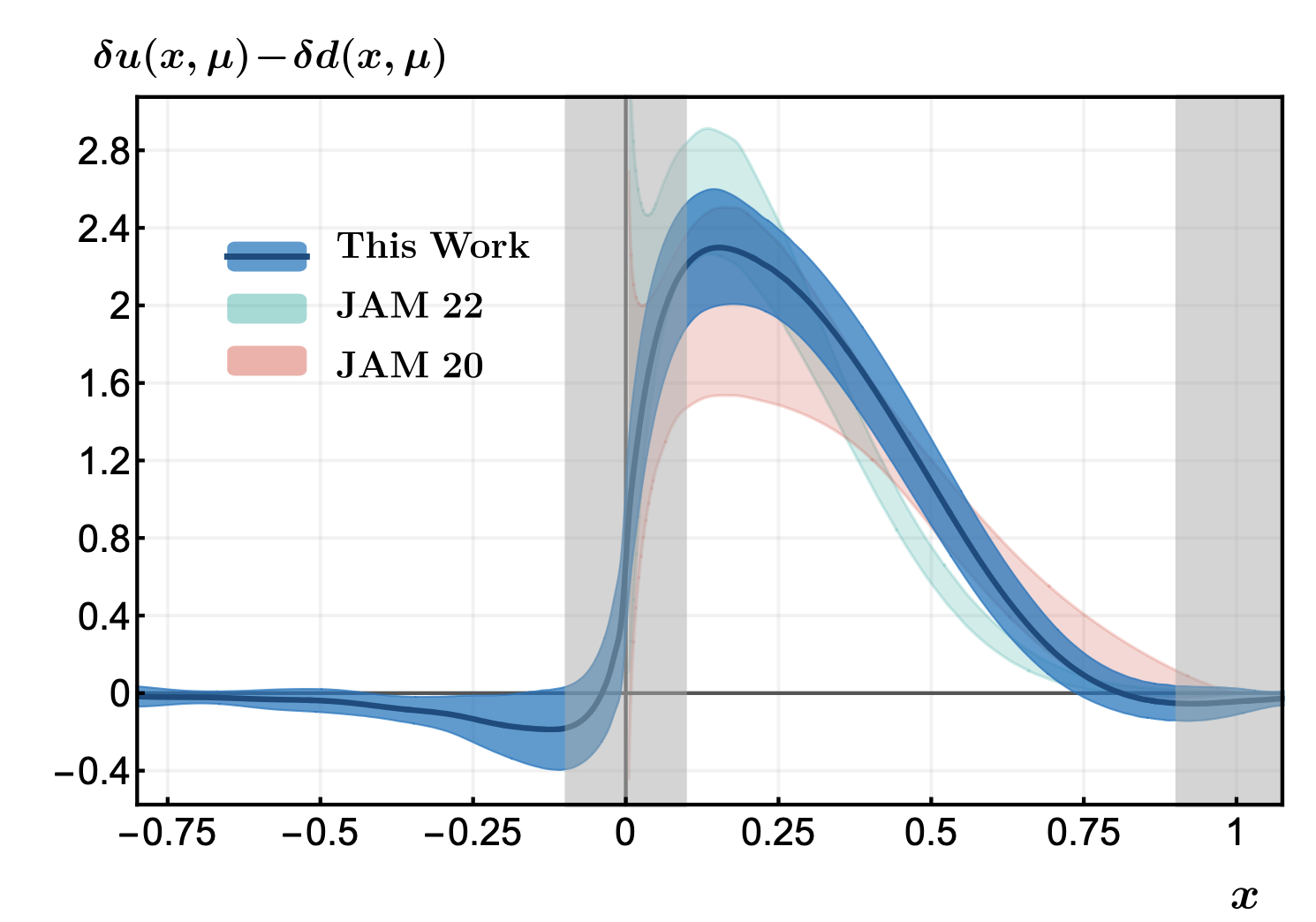}
\end{minipage}
\begin{minipage}{0.3\linewidth}
\hspace*{-1.cm}\includegraphics[scale=0.2]{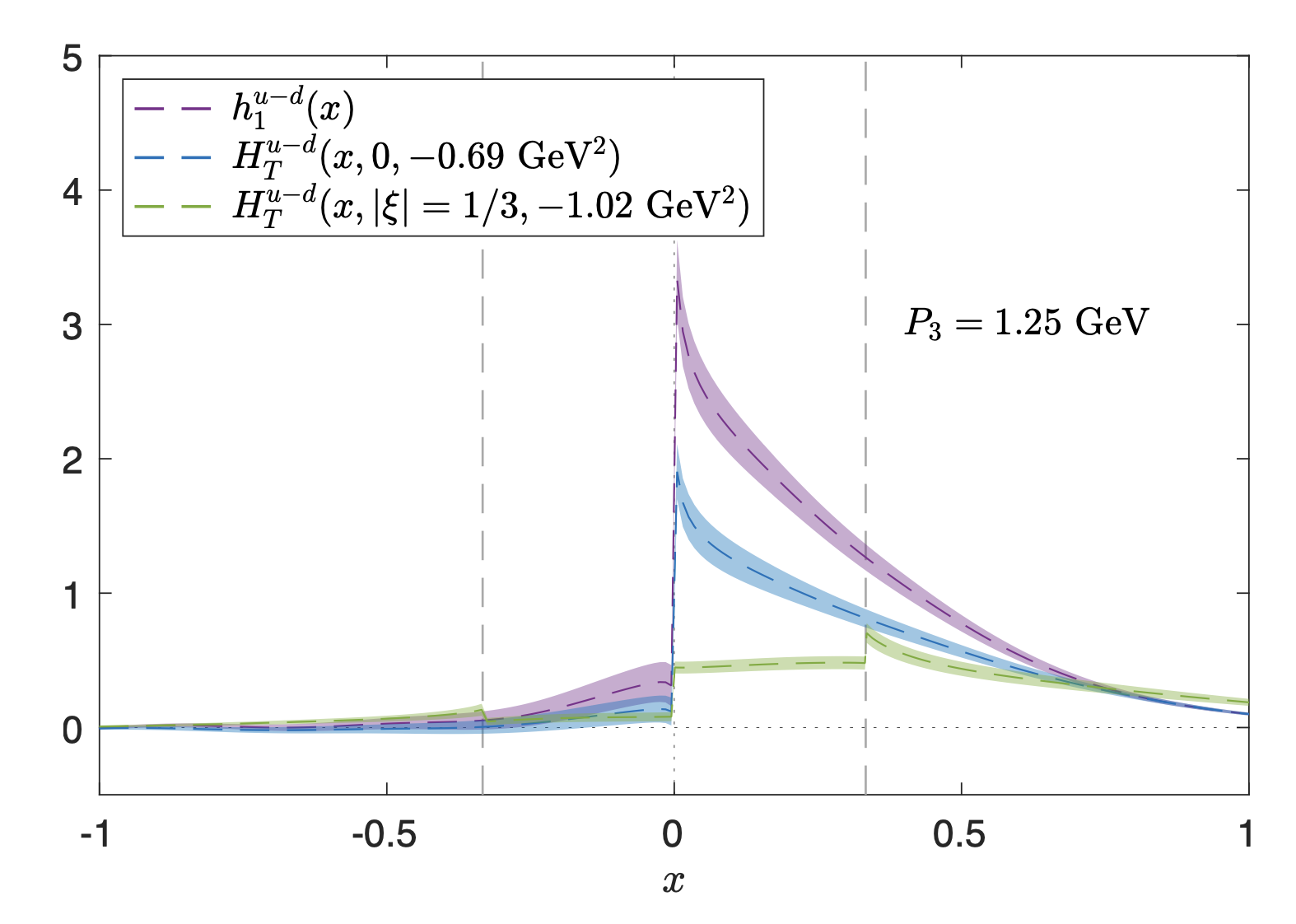}
\end{minipage}
\caption{Left: The isovector helicity PDF from Ref.~\cite{Bringewatt:2020ixn}.  The JAM17 results are shown
with red bands, while the blue bands show the resulting fits when using both lattice and experimental data. The yellow bands show the lattice QCD data from Ref.~\cite{Alexandrou:2018pbm}. Middle: The isovector transversity PDF from Ref.~\cite{LatticeParton:2022xsd}. Left: The transversity GPD for skewness 1/3 from Ref.~\cite{Alexandrou:2021bbo}. }\label{fig:PDFs}
\end{figure}
Spatial correlators can be readily  computed in lattice QCD yielding the so-called quasi-PDF  $\tilde{F}_\Gamma(x,P_3,\mu)$
\begin{equation}
  \tilde{F}_\Gamma(x,P_3,\mu) =2P_3
  \int_{-\infty}^{\infty}\frac{dz}{4\pi}e^{-ixP_3z}\,\langle P_3\vert\,\overline{\psi}(0)\, \Gamma W(0,z)\, \psi(z)|\,P_3\rangle |_{\mu}\quad ,
\end{equation}
 renormalized non-perturbatively.
The quasi-PDF is then matched using a kernel $C$ to determine  the PDF ${F}_\Gamma(y,\mu) $
\be
\tilde{F}_\Gamma(x,P^z,\mu)=\int_{-1}^1
\frac{dy}{|y|} \, C\left(\frac{x}{y},\frac{\mu}{yP^z}\right)\, {F}_\Gamma(y,\mu) +{\cal{O}}\left(\frac{\Lambda^2_{\rm QCD}}{(xP^z)^2},\frac{\Lambda^2_{\rm QCD}}{((1-x)P^z)^2}\right).
\ee
In the past few years, PDFs and  GPDs are being computed by  several groups following this approach. For reviews, see Refs.~\cite{Lin:2017snn,Constantinou:2020hdm}. As an example, we show results on the isovector helicity and transversity PDFs obtained by ETMC~\cite{Alexandrou:2018pbm} and the Lattice Parton Collaboration (LPC)~\cite{LatticeParton:2022xsd}, respectively,  in Fig.~\ref{fig:PDFs}. These results are already providing constrains to phenomenological extractions, as indicated by comparing the precision obtained for the helicity PDF when lattice QCD data are used as input to those without. We also show the isovector transversity GPD at skewness $\xi=1/3$ that demonstrates the feasibility to determine directly GPDs in lattice QCD within LaMET. A similar approach can be applied to compute TMDs. \vspace*{-0.3cm}

\section{Conclusions}\label{sec:Conclusions}
The  lattice framework provides the {\it ab initio} approach to study nucleon structure with improvable statistical and systematic errors. Theoretical and algorithmic progress together with larger computational resources are enabling lattice QCD  simulations to be performed using physical values of the light, strange and charm quarks at several lattice spacings ranging from about 0.1~fm to 0.5~fm and large enough volumes. With such simulations, lattice QCD can deliver  precision results for a number of key quantities, such as  nucleon charges, form factors and second Mellin moments that provide insights for its structure.  New theoretical developments are also extending the range of quantities that can be extracted within lattice QCD. Examples discussed are the direct computation of PDFs and GPDs providing valuable input to phenomenology and experiment. 

 \vspace*{0.3cm}

\noindent
{\bf Acknowledgements}\vspace*{0.2cm}\\
I would like to thank the members of ETMC for their contributions to the realization of the results presented. I also acknowledge partial support from the  project 3D-nucleon funded by the European Regional Development Fund and the Republic of Cyprus through the Cyprus Research and Innovation Foundation under contract number EXCELLENCE/0421/0043 and the European Joint Doctorate project AQTIVATE funded by the European Commission under the Marie Sklodowska-Curie Doctoral Networks action and Grant Agreement No 101072344. Furthermore, I gratefully acknowledge PRACE for awarding access to HAWK at HLRS within the project with Id Acid 4886, and the Swiss National Supercomputing Centre (CSCS) and the EuroHPC Joint Undertaking for awarding this project access to the LUMI supercomputer, owned by the EuroHPC Joint Undertaking, hosted by CSC (Finland) and the LUMI consortium through the Chronos programme under project IDs CH17-CSCS-CYP and CH21-CSCS-UNIBE. 

\bibliographystyle{JHEP}
\bibliography{ref}


\end{document}